\documentclass[aps,prl,amsmath,amssymb,twocolumn,reprint,showpacs,noshowkeys]{revtex4-1}
\pdfoutput=1 
\usepackage[latin1]{inputenc}
\usepackage{graphicx}
\usepackage{color}
\usepackage{hyperref}
\begin{document}

\title{Fiber-coupled single ion as an efficient quantum light source}

\author{Alex Wilson} 
\author{Hiroki Takahashi}
\author{Andrew Riley-Watson}
\author{Fedja Oru\v{c}evi\'{c}} 
\author{Peter Blythe}
\author{Anders Mortensen}
\author{Daniel R.\ Crick}
\author{Nicolas Seymour-Smith}
\author{Elisabeth Brama}
\author{Matthias Keller}
\author{Wolfgang Lange}
\email{W.Lange@sussex.ac.uk}
\affiliation{Department of Physics and Astronomy, University of Sussex,
Brighton, BN1 9QH, United Kingdom}
\date{January 26,2011}

\begin{abstract} 
We have realized a compact system to efficiently couple the fluorescent light
emitted by a single trapped ion to two opposing optical fibers. 
The fibers are tightly integrated in the center electrodes of a miniature endcap
trap. They capture light from the ion with a numerical aperture of 0.34 each,
corresponding to 6\% of the solid angle in total. The high collection efficiency 
and high signal-to-background ratio make the setup an ideal quantum light source.
We have observed strong antibunching of the photons emitted from
the two fibers. The system has a range of applications from single-ion
state detection in quantum information processing to strong coupling
cavity-QED with ions. 
\end{abstract}


\pacs{37.10.Ty, 42.81.-i, 32.50.+d, 42.50.-p, 03.67.Lx}
\keywords{photon antibunching, single atom, optical fiber, ion trap,
quantum computation, fluorescence}

\maketitle

The interaction of light and single atomic particles has evolved from 
a subject of fundamental interest to a unique tool for modern quantum
technology. The fluorescent radiation emitted by single atoms has pronounced 
quantum properties, as demonstrated by the second order correlation 
function \cite{Kimbl77}. While early experiments provided little 
control over individual atomic particles, advances in trapping and 
cooling of atoms and ions have led to a range of applications, most 
notably atom chips and quantum information processing in strings
of ions.

An important figure of merit of these systems is the efficiency with which radiation 
emitted by a single particle is detected. In atom chips, this is essential for the 
ability to detect the presence of atoms \cite{Horak03}. In ion traps, more efficient 
collection of fluorescent light speeds up quantum state discrimination, as required 
for the read-out of a quantum register \cite{Myers08}.  

As a prerequisite for efficient detection, a high numerical aperture (NA)
system must be employed for collecting radiation. The ideal collection efficiency 
would be achieved if photons from the full solid angle were captured. Recently, 
64\% of the emission of freely falling atoms was collected by
combining two mirrors \cite{Bondo06}. There are proposals to approach 100\% 
capture efficiency by surrounding a single atom with a parabolic mirror \cite{Lindl07}. 
In experiments with single ions, the need for trap electrodes and laser access 
requires a more open structure. Collection efficiencies of 10\% have been reached
with a spherical mirror integrated with a linear Paul trap \cite{Shu10}.
Another approach is to surround the particle with a pair of lenses with large
NA \cite{Tey09}. 

In systems with large optical elements, the geometry of trap and vacuum chamber
poses limitations for light detection. A novel way of capturing fluorescence 
is provided by optical fibers. While the diameter of the fiber core is only on 
the order of 100~$\mu$m, the fiber ends can be brought in close proximity to 
the fluorescing particle to maximize the numerical aperture. Integrated fibers 
have been used successfully to detect the presence of atoms on a chip, 
either in absorption \cite{Quint04} or fluorescence \cite{Wilzb09}. Even 
tighter integration of fluorescence collection on atom chips has been proposed
by using optical waveguides \cite{Kohne09}.

\begin{figure}
  \centering
  \begin{minipage}[t]{0.34 \linewidth}
  {\sf a)}\hfill
  \vspace{2pt}
  \includegraphics[width=\linewidth]{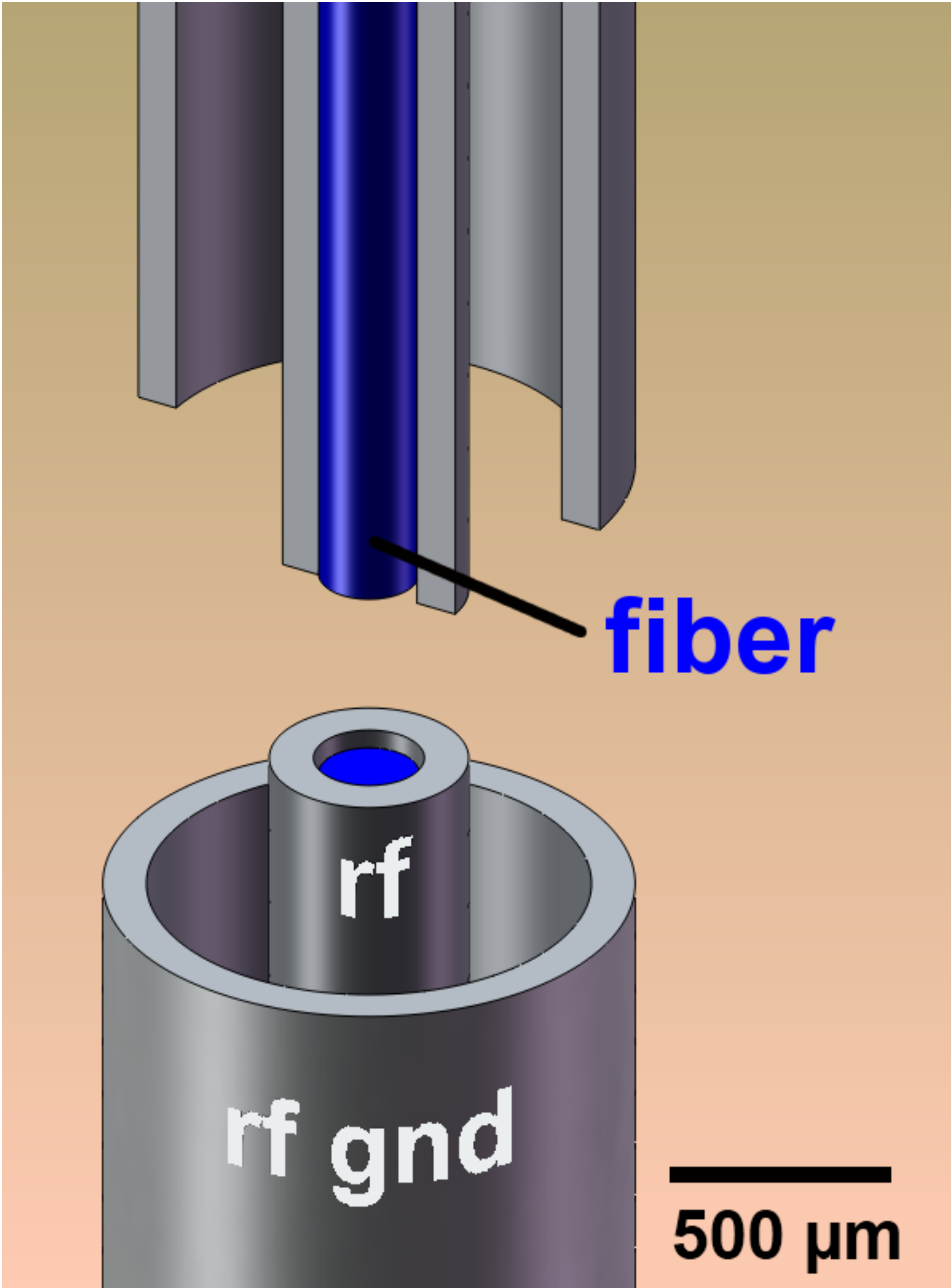} 
  \end{minipage} \hfill
  \begin{minipage}[t]{0.64 \linewidth}
  \hbox to \linewidth{\hspace{7mm}{\sf b)}} 
  \includegraphics[width=\linewidth]{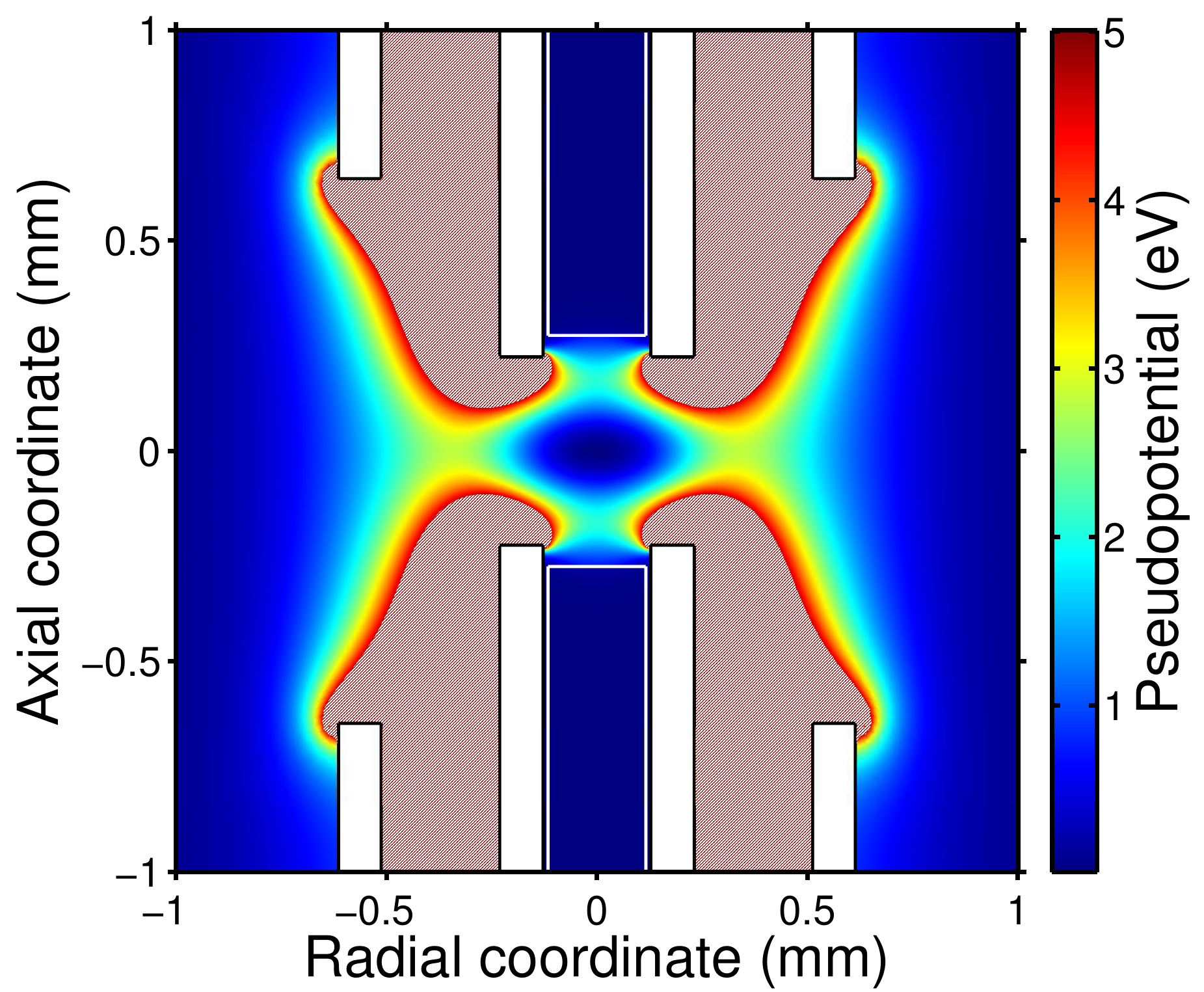}
  \end{minipage} 
  \caption{(color online). (a) Endcap trap with integrated optical fibers. Half of the 
    upper electrode structure is cut away to reveal the upper fiber. 
   (b) Cross-section of the pseudopotential of the trap with fiber, 
   obtained from a finite element calculation by averaging the potential 
   over one rf-cycle. 
  } 
  \label{fig:trapscheme}
\end{figure}

Using optical fibers for collecting fluorescence from trapped \emph{ions} is 
complicated by the fact that the trapping potential is adversely affected by 
the presence of dielectrics. Proper shielding and tight integration of the 
fiber in the trapping structure is required \cite{VanDe10, *Brady10}. We have 
realized a novel system for combining optical fibers with an rf ion-trap. 
It fulfills the requirements of close proximity of the fiber ends to the ion, 
but at the same time negligible distortion of the trapping field. Our trap 
design is of the endcap type \cite{Schra93}, in which a single ion 
is stored between the ends of two opposing metal rods connected to an rf-source. 
Additional hollow coaxial electrodes at rf-ground, surrounding the central 
electrodes, increase the depth of the trapping 
potential. The cylindrical symmetry of the trap provides a natural way of 
combining it with optical fibers. By replacing the central rf-rods with 
hollow tubes, optical fibers can be inserted in both electrodes 
[Fig.~\ref{fig:trapscheme}(a)]. The fibers are not flush with the end of the 
tube, but retracted by 50~$\mu$m, so that rf-shielding inside the tube
makes the overlap of the dielectric fibers and the rf-field negligible. 
In addition, the presence of the fibers does not change the cylindrical 
symmetry of the trap, limiting distortion of the trapping field. 

The outer diameter of the central electrodes is 458~$\mu$m, the inner
diameter 254~$\mu$m, and the vertical distance between the electrodes is
446~$\mu$m. The optical fiber in our experiment is a Thorlabs BFH48-200
multimode fiber with a core diameter of 200~$\mu$m and a NA of 0.48. The 
hard cladding diameter is 230~$\mu$m so that the fiber can be
run down the center electrodes with the coating stripped. After inserting 
the fibers from the back of the electrodes and positioning their end-facets,
we fixed them by slightly squeezing the rear of the rf-electrodes. 
At an ion-fiber separation of 275~$\mu$m, the effective NA of the fiber 
core is 0.34. At this setting, the two fibers combined capture 6\% of
the full solid angle, only limited by the geometry. The capture efficiency of our
setup is approximately twice of what has been demonstrated so far for fibers
integrated with ion traps \cite{VanDe10, *Brady10}. Using the full numerical
aperture of the fiber, a capture efficiency of 12.3\% can be achieved at an 
ion-fiber separation of 183~$\mu$m. 

Calcium ions are loaded from an effusive oven mounted to the side of the trap.
To reduce coating of the trap electrodes with calcium, the atomic beam is
collimated by means of a tube of inner diameter 250~$\mu$m, 2.45~mm away from
the center of the trap. The atoms are photoionized using a resonant laser at
423~nm and a second stage at 375~nm for the actual ionization \cite{Gulde01}. Once captured in
the trapping potential, a single ion is stored for several hours. Compared to 
an endcap trap with solid central electrodes, tubular electrodes generate a 
pseudopotential with a 25\% lower trap depth [Fig~\ref{fig:trapscheme}(b)]. For 
an rf-amplitude of 200~V, the calculated potential barriers are 2.8~eV in 
the radial and 2.1~eV in the axial direction. The secular frequencies were 
measured to be $\omega_r$=$(2\pi)1.9$~MHz and $\omega_z$=$(2\pi)3.8$~MHz for a 
drive frequency of 14.9~MHz.

\begin{figure}
  \centering
  \includegraphics[width=0.7 \linewidth]{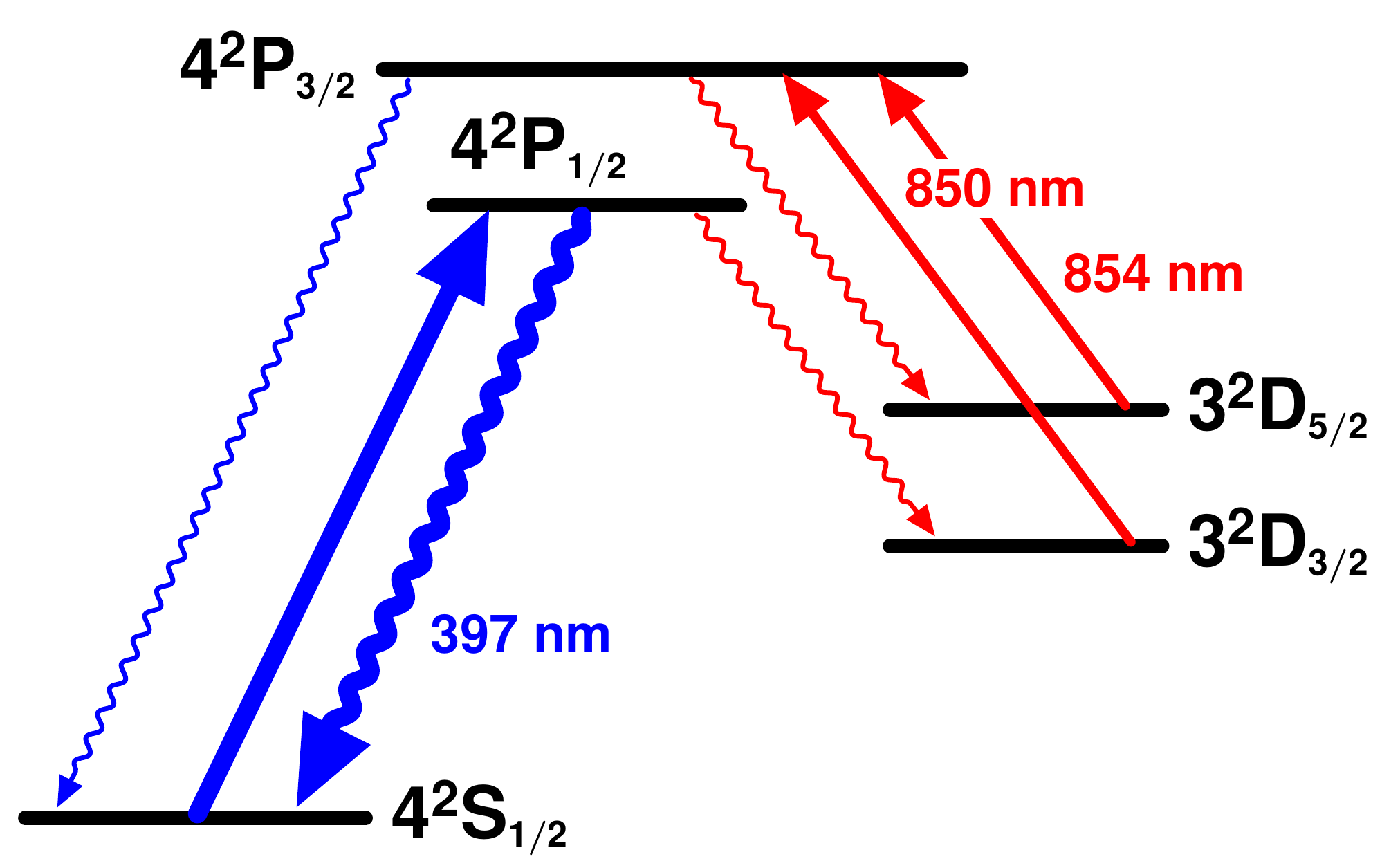}
  \caption{(color online). Scheme of the relevant transitions in 
  ${}^{40}\mathrm{Ca}^{+}$. We probe the quantum properties of
  light scattered on the resonance transition at 397~nm (bold
  arrows). Population trapping due to decay to the D-states is
  avoided by repumping via the upper P-state (thin arrows).}
  \label{fig:levelscheme}
\end{figure}

With a view to applications in spectroscopy, quantum information 
processing and cavity-QED, the localization of the 
ion in the trap is of particular importance. As a first step, 
the ion is laser-cooled 
on the $S_{1/2}\rightarrow P_{1/2}$ transition with a wavelength 
of $\lambda$=397~nm. Lasers with a power of several $\mu$W are injected 
from the side under different angles, red-detuned by roughly half the 
linewidth $\Gamma$=$(2\pi)22.3$~MHz. To avoid optical pumping 
to the $D_{\text{3/2}}$ and the $D_{\text{5/2}}$ level, we apply 
lasers at 850~nm and and 854~nm, returning the ion to the ground 
state via the $P_{\text{3/2}}$-level [Fig~\ref{fig:levelscheme}].

The ion may also undergo a periodic oscillation, driven by the 
rf-trapping field (micromotion), if dc electric stray fields push
the ion off the central node of the rf-field. In order 
to minimize micromotion, we compensate stray fields by applying 
voltages to a set of compensation electrodes. The atomic beam collimator 
serves 
as the electrode compensating dc stray fields in the direction of 
the oven. Horizontal dc-fields orthogonal to the 
oven are compensated via a small wire mounted to the side,
while a vertical dc offset-field can be applied via the rf-ground 
electrodes. With the help of these electrodes, we position the ion at 
the rf-center of the trap, where its micromotion is minimal. The 
required dc-voltages are determined by probing the rf-modulation of 
the fluo\-re\-scence intensity using UV pump-beams in three 
non-collinear directions. 

The ion's localization that can be achieved depends on the sensitivity
with which we detect micromotion. 
The sensitivity to Doppler modulation in our experiment is  
$0.016\Gamma/\sqrt{\mathrm{Hz}}$. For an acquisition time of 4~s, we
can detect an axial displacement of $0.006\Gamma\lambda/\omega_z \approx 0.04\lambda$. 
Therefore, by compensating stray fields and nulling the micromotion, we localize the ion 
in a region smaller than the wavelength (Lamb-Dicke regime), a fact which
is essential for applications in cavity-QED.

The presence of the fiber facets close to the ion might potentially
lead to stray electric fields at the position of the ion due to the accumulation 
of charges on dielectric surfaces \cite{Harla10,Kim10}. If left 
uncompensated, the ion would be pushed off the 
rf-node and hence become subject to micromotion, 
resulting in line broadening and reduced coupling to light. Charges might be
created by direct laser-illumination of surfaces or in the photoionization of 
atomic calcium. We minimize these effects by reducing the beam waists and recessing
the end of the fibers as described above. Another source of stray fields
are contact potentials in sections of the electrodes partially coated with
calcium.

As a sensitive probe for the presence of stray fields, we have utilized the 
trapped ion itself. Automatically nulling the micromotion at intervals of 2 
minutes, we tracked the compensation voltages over time after loading the 
trap. Using a model of the trap fields obtained from a 3D finite element 
calculation, the compensation voltages are converted to an electric 
field at the center of the trap which must be equal and diametrically 
opposed to the instantaneous stray field. The sensitivity of the
measurement was $60\mathrm{~m\!V}/(\mathrm{cm}\sqrt{\mathrm{Hz}})$. Our data showed 
stationary conditions during normal operation of the trap. Immediately 
after loading, stray fields on the order of 1~V/cm appear,
but decay exponentially at a rate of $5\times 10^{-4} \mathrm{s}^{-1}$
on average.
Since we measure all three spatial components of the electric field, 
we can locate where it originates from.
We find that the sources lie in a small azimuthal segment on the edge 
of the central rf-electrodes, on the side facing the oven. There is no
evidence of fields from the direction of the fibers. This is a strong 
indication that charging of the fiber end-facets does not play a 
major role, but that patch potentials on the electrodes are responsible.

In order to distinguish between the effects of laser illumination
and the atomic beam, we switched the photoionization lasers
and oven on individually, then observed the effect on an ion already 
present in the trap. The atomic beam on its own did not
affect the micromotion of the ion.
Laser illumination at 423~nm and 375~nm alone lead to a change 
in compensation voltages, but it was a factor of 30 smaller than 
after loading the trap. This resilience to charging is further evidence for
the compatibility of optical fibers with our trap setup. 
 
With the trapped ion centered between the fiber ends and the fluorescent
light captured and guided to the detectors by optical fibers, there is 
no need for additional optics or any optical alignment, making the setup 
very easy to maintain. Both fibers are connected to
atmosphere via vacuum feedthroughs (core diameter 400~$\mu$m) which in 
turn are linked to photomultiplier tubes (PMT, Hamamatsu H5773) by another 
optical fiber (core diameter 600~$\mu$m). Total transmission is approximately 
80\%. We have investigated the quantum properties of light emitted on the 
$P_{1/2}\rightarrow S_{1/2}$ transition [Fig.~\ref{fig:levelscheme}].
The small dimensions of the trap require special care to prevent stray light   
from entering the optical fibers and reducing the signal-to-background ratio
(SBR)\@. In order to minimize scattered light, we spatially filter the principal
397~nm beam and focus it with a diffraction limited lens (NA=0.25). In this
way, we strongly suppress the background count rate in both channels. 

\begin{figure}
  \centering
  \includegraphics[width=0.75 \linewidth, angle=0]{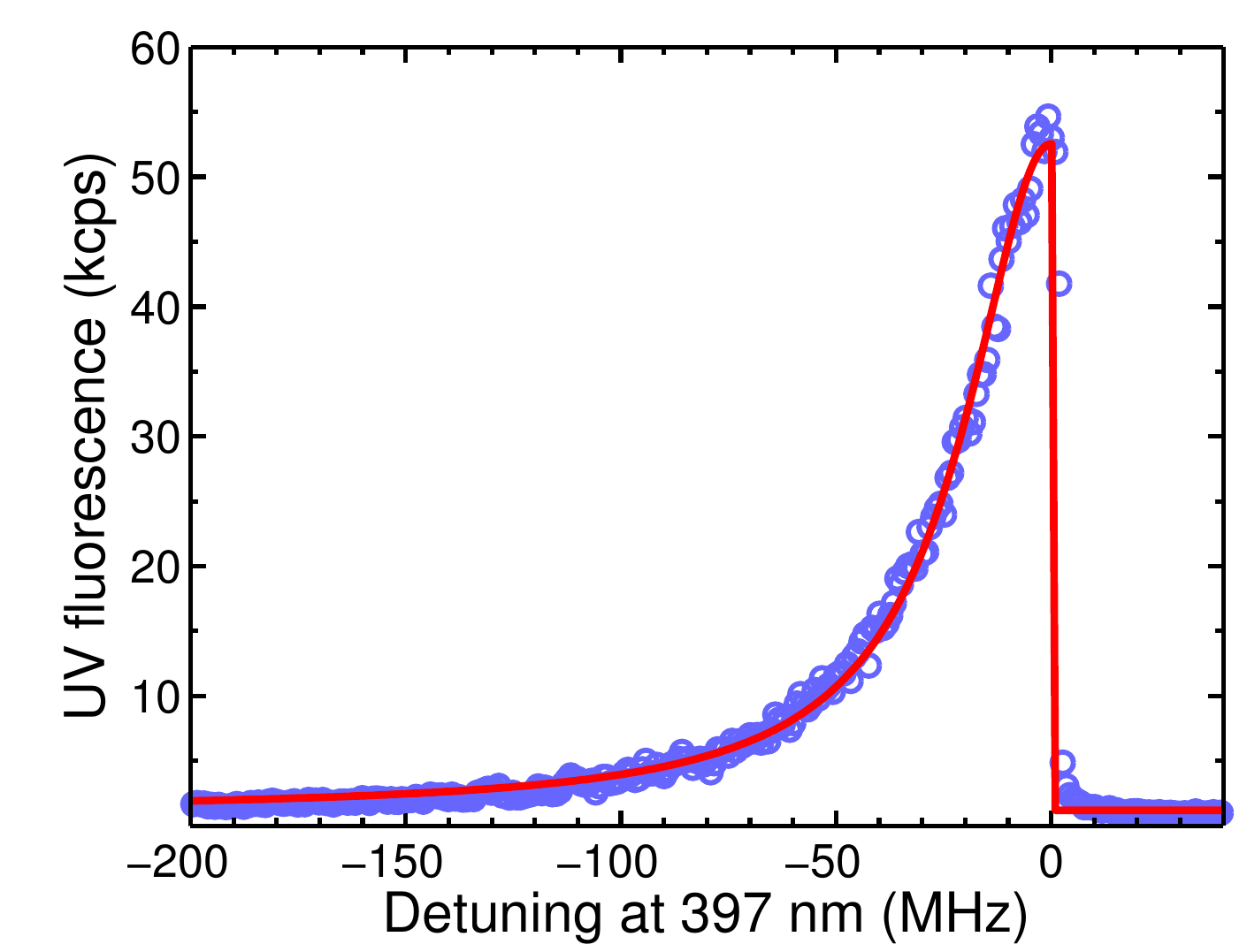}
  \caption{(color online). Fluorescence spectrum of a single ion in the fiber-coupled endcap
  trap. Excitation and repumping beams at a power of 0.39~$\mu$W 
  (equivalent to an intensity of $1.3I_s)$ and 0.5~mW,   
  respectively, are injected from the side of the trap. The combined 
  fluorescence of the two fibers is plotted and no background subtracted. 
  It is fit by a Lorentzian with a half width of 23.8~MHz (solid line).}
  \label{fig:line}
\end{figure}

Spectroscopy with fiber-based fluorescence detection is demonstrated by 
scanning the 397~nm laser over the red-detuned half of the resonance 
line of the calcium ion, where laser-cooling occurs. Combining the signal 
from both fibers, we measur a peak fluorescence count rate of 36k~counts 
per second (cps) at the saturation intensity ($I_s$) against a background of 740~cps
due to stray light and PMT dark counts. This corresponds to a SBR of 49. 
At slightly higher laser intensities, count rates up to 55~kpcs are 
observed [Fig~\ref{fig:line}]. Thus, the combination of endcap-trap and 
optical fibers provides excellent conditions for spectroscopy of a 
single ion.

It is well known that the fluorescent light of a single atomic emitter has non-classical
characteristics. They can be probed via the photon statistics 
which shows perfect antibunching \cite{Carmi76}. 
Antibunched fluorescence from a single atom was first
demonstrated in an atomic beam \cite{Kimbl77} and later with ions
\cite{Diedr87}. The fact that the second order correlation function is
close to zero at times smaller than the relaxation time of the transition 
can be interpreted as the emission of photons one by one, separated by
the required atomic repumping. This makes a single driven ion an 
efficient source of single photons, even though emission occurs at 
random times. 

The fiber-coupled endcap trap allows us to directly investigate the quantum 
properties of the fluorescent light from a single ion. Each of the two fibers 
delivers an antibunched stream of photons. Even more importantly, the
arrival times of photons in the two fibers are anticorrelated, as they originate
from the same single ion. Our setup is an ideal, miniaturized version of the
Hanbury-Brown Twiss experiment. Instead of generating two anticorrelated photon
streams with the help of a beam splitter down the optical path, we use the photons
captured by the two optical fibers directly.  

\begin{figure}[t]
  \centering
  \includegraphics[width=0.8 \linewidth, angle=0, trim = 0mm 0mm 0mm 4mm, clip=true]{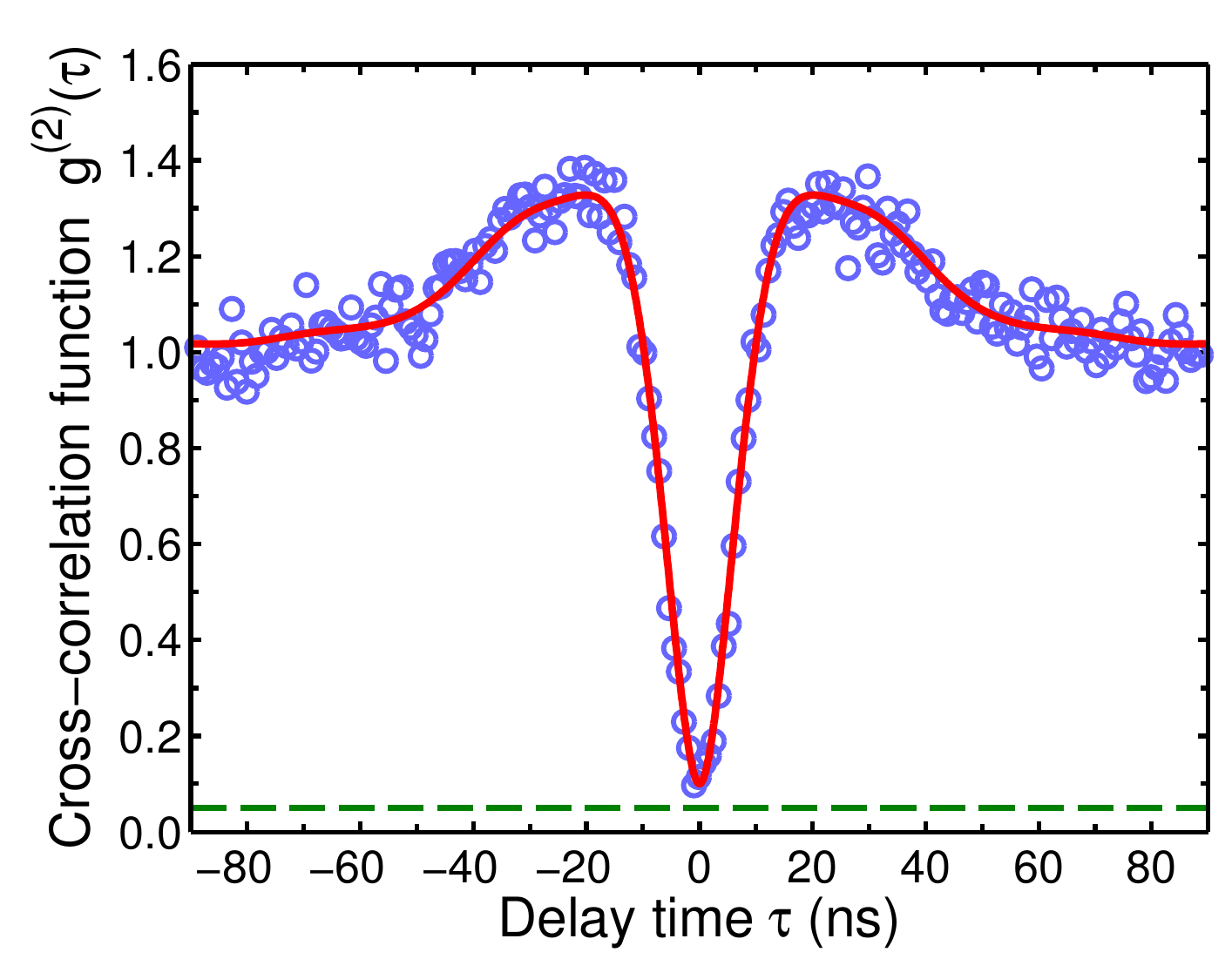}
  \caption{(color online). Normalized cross-correlation signal $g^{(2)}(\tau)$ of photon arrival
   times at the end of the two fibers. The dashed line indicates the expected 
   background level due to stray light and dark counts, which have not been 
   subtracted. The excellent agreement with the superimposed numerical 
   solution of the master equation demonstrates the distinct quantum 
   character of the emitted radiation.}
  \label{fig:g2}
\end{figure}

We measure second order correlations between photons in the two fibers by
sending the PMT-output to a time-to-digital converter (TDC, FAST 7072T)\@. The
device has two separate start and stop channels, which we set up so that
photons from each fiber can trigger a correlation measurement, to be stopped
by a photon from the other fiber. In this way, we are able to evaluate
cross-correlations of all registered photon pulses. A delay of  
200~ns in each stop-channel gives us access to positive and negative correlation
times. Finally, the two TDC traces are aligned at $\tau=0$ and added.
Figure~\ref{fig:g2} shows a correlation measurement acquired in 40 minutes, with 
the driving laser red-detuned by 6~MHz and at a power of 0.16~$\mu$W. The 
signal-to-background ratio in the two fiber channels is
\textit{SBR}$_1=75$  and  \textit{SBR}$_2=26$, respectively. The difference 
between them is due to a slight angle of the pump beam, scattering different 
amounts of light into the upper and lower fibers. The background contribution 
leads to an offset of 0.05 in the normalized correlation function $g^{(2)}(\tau)$ 
[see Fig.~\ref{fig:g2}]. After subtracting this offset, we obtain 
$g^{(2)}(0)=0.05\pm0.04$, an important figure of merit attesting to the quality 
of our system as a single-photon source. The measured correlation function is in 
excellent agreement with the solution of the full master equation of the calcium 
ion, obtained numerically and also shown in Fig~\ref{fig:g2}.

The geometry of a single ion between two opposing optical fibers has
important applications beyond efficient state-detection, spectroscopy and the
generation of non-classical radiation.
By furnishing one or both end-facets of the fibers with a concave surface and
providing them with a high-reflectivity coating, an optical resonator can be
established, known as a fiber-cavity \cite{Colom07}. A
radius of curvature around 100~$\mu$m is readily achieved using ablation with a
high-power laser. The small separation between the end-facets and the small
radius of curvature lead to a very small volume of the cavity modes and hence to
strong coupling between the electromagnetic field and an ion inside the cavity.
The high stability and long trapping times achieved with single ions make it
very attractive to strongly couple them to the field of a fiber-cavity. The
endcap-trap geometry implemented here is an ideal way of avoiding the 
fundamental problem of dielectric materials distorting the rf-trapping field. 
Our results show that reliable storage of a single ion is possible in the 
proximity of optical fibers. Finally, note that the architecture of two 
optical fibers coupled to an ion is ideally suited for the realization of an 
ion-based quantum repeater \cite{Sango09}.

In summary, we have captured the fluorescence of a single ion trapped in an
endcap trap, using a pair of optical fibers integrated inside the two central
rf-electrodes. Connecting each fiber to a photomultiplier tube provided us with
a highly efficient detection system which is easy to maintain, as no optical
adjustments are required. The solid angle subtended by the two fibers is 6\% of 
$4\pi$, making it the most efficient fiber detection system implemented for ions
so far. We have analyzed the photon statistics of the light emitted by a single
ion and have observed distinct non-classical properties of the radiation, again
without any additional optical elements in the setup. The high signal-to-background 
ratio of 49 leads to a strong antibunching signal. The geometry of the system 
is identical to that required for
coupling a single ion to an optical fiber-cavity and our results show
unequivocally that it is technically feasible to provide stable trapping of a
single ion in such a setup. 

We gratefully acknowledge support from the European Commission (Marie Curie Excellence 
Grant MEXT-CT-2005-025703) and the EPSRC (EP/D061296/1).


\begin{thebibliography}{20}%
\makeatletter
\providecommand \@ifxundefined [1]{%
 \@ifx{#1\undefined}
}%
\providecommand \@ifnum [1]{%
 \ifnum #1\expandafter \@firstoftwo
 \else \expandafter \@secondoftwo
 \fi
}%
\providecommand \@ifx [1]{%
 \ifx #1\expandafter \@firstoftwo
 \else \expandafter \@secondoftwo
 \fi
}%
\providecommand \natexlab [1]{#1}%
\providecommand \enquote  [1]{``#1''}%
\providecommand \bibnamefont  [1]{#1}%
\providecommand \bibfnamefont [1]{#1}%
\providecommand \citenamefont [1]{#1}%
\providecommand \href@noop [0]{\@secondoftwo}%
\providecommand \href [0]{\begingroup \@sanitize@url \@href}%
\providecommand \@href[1]{\@@startlink{#1}\@@href}%
\providecommand \@@href[1]{\endgroup#1\@@endlink}%
\providecommand \@sanitize@url [0]{\catcode `\\12\catcode `\$12\catcode
  `\&12\catcode `\#12\catcode `\^12\catcode `\_12\catcode `\%12\relax}%
\providecommand \@@startlink[1]{}%
\providecommand \@@endlink[0]{}%
\providecommand \url  [0]{\begingroup\@sanitize@url \@url }%
\providecommand \@url [1]{\endgroup\@href {#1}{\urlprefix }}%
\providecommand \urlprefix  [0]{URL }%
\providecommand \Eprint [0]{\href }%
\providecommand \doibase [0]{http://dx.doi.org/}%
\providecommand \selectlanguage [0]{\@gobble}%
\providecommand \bibinfo  [0]{\@secondoftwo}%
\providecommand \bibfield  [0]{\@secondoftwo}%
\providecommand \translation [1]{[#1]}%
\providecommand \BibitemOpen [0]{}%
\providecommand \bibitemStop [0]{}%
\providecommand \bibitemNoStop [0]{.\EOS\space}%
\providecommand \EOS [0]{\spacefactor3000\relax}%
\providecommand \BibitemShut  [1]{\csname bibitem#1\endcsname}%
\let\auto@bib@innerbib\@empty
\bibitem [{\citenamefont {{Kimble}}\ \emph {et~al.}(1977)\citenamefont
  {{Kimble}}, \citenamefont {{Dagenais}},\ and\ \citenamefont
  {{Mandel}}}]{Kimbl77}%
  \BibitemOpen
  \bibfield  {author} {\bibinfo {author} {\bibfnamefont {H.~J.}\ \bibnamefont
  {{Kimble}}}, \bibinfo {author} {\bibfnamefont {M.}~\bibnamefont
  {{Dagenais}}}, \ and\ \bibinfo {author} {\bibfnamefont {L.}~\bibnamefont
  {{Mandel}}},\ }\href {http://dx.doi.org/10.1103/PhysRevLett.39.691}
  {\bibfield  {journal} {\bibinfo  {journal} {Phys. Rev. Lett.}\ }\textbf
  {\bibinfo {volume} {39}},\ \bibinfo {pages} {691} (\bibinfo {year}
  {1977})}\BibitemShut {NoStop}%
\bibitem [{\citenamefont {{Horak}}\ \emph {et~al.}(2003)\citenamefont
  {{Horak}}, \citenamefont {{Klappauf}}, \citenamefont {{Haase}}, \citenamefont
  {{Folman}}, \citenamefont {{Schmiedmayer}}, \citenamefont {{Domokos}},\ and\
  \citenamefont {{Hinds}}}]{Horak03}%
  \BibitemOpen
  \bibfield  {author} {\bibinfo {author} {\bibfnamefont {P.}~\bibnamefont
  {{Horak}}}, \bibinfo {author} {\bibfnamefont {B.~G.}\ \bibnamefont
  {{Klappauf}}}, \bibinfo {author} {\bibfnamefont {A.}~\bibnamefont {{Haase}}},
  \bibinfo {author} {\bibfnamefont {R.}~\bibnamefont {{Folman}}}, \bibinfo
  {author} {\bibfnamefont {J.}~\bibnamefont {{Schmiedmayer}}}, \bibinfo
  {author} {\bibfnamefont {P.}~\bibnamefont {{Domokos}}}, \ and\ \bibinfo
  {author} {\bibfnamefont {E.~A.}\ \bibnamefont {{Hinds}}},\ }\href
  {http://dx.doi.org/10.1103/PhysRevA.67.043806} {\bibfield  {journal}
  {\bibinfo  {journal} {\pra}\ }\textbf {\bibinfo {volume} {67}},\ \bibinfo
  {pages} {043806} (\bibinfo {year} {2003})}\BibitemShut {NoStop}%
\bibitem [{\citenamefont {{Myerson}}\ \emph {et~al.}(2008)\citenamefont
  {{Myerson}}, \citenamefont {{Szwer}}, \citenamefont {{Webster}},
  \citenamefont {{Allcock}}, \citenamefont {{Curtis}}, \citenamefont {{Imreh}},
  \citenamefont {{Sherman}}, \citenamefont {{Stacey}}, \citenamefont
  {{Steane}},\ and\ \citenamefont {{Lucas}}}]{Myers08}%
  \BibitemOpen
  \bibfield  {author} {\bibinfo {author} {\bibfnamefont {A.~H.}\ \bibnamefont
  {{Myerson}}}, \bibinfo {author} {\bibfnamefont {D.~J.}\ \bibnamefont
  {{Szwer}}}, \bibinfo {author} {\bibfnamefont {S.~C.}\ \bibnamefont
  {{Webster}}}, \bibinfo {author} {\bibfnamefont {D.~T.~C.}\ \bibnamefont
  {{Allcock}}}, \bibinfo {author} {\bibfnamefont {M.~J.}\ \bibnamefont
  {{Curtis}}}, \bibinfo {author} {\bibfnamefont {G.}~\bibnamefont {{Imreh}}},
  \bibinfo {author} {\bibfnamefont {J.~A.}\ \bibnamefont {{Sherman}}}, \bibinfo
  {author} {\bibfnamefont {D.~N.}\ \bibnamefont {{Stacey}}}, \bibinfo {author}
  {\bibfnamefont {A.~M.}\ \bibnamefont {{Steane}}}, \ and\ \bibinfo {author}
  {\bibfnamefont {D.~M.}\ \bibnamefont {{Lucas}}},\ }\href
  {http://dx.doi.org/10.1103/PhysRevLett.100.200502} {\bibfield  {journal}
  {\bibinfo  {journal} {Phys. Rev. Lett.}\ }\textbf {\bibinfo {volume} {100}},\
  \bibinfo {pages} {200502} (\bibinfo {year} {2008})}\BibitemShut {NoStop}%
\bibitem [{\citenamefont {{Bondo}}\ \emph {et~al.}(2006)\citenamefont
  {{Bondo}}, \citenamefont {{Hennrich}}, \citenamefont {{Legero}},
  \citenamefont {{Rempe}},\ and\ \citenamefont {{Kuhn}}}]{Bondo06}%
  \BibitemOpen
  \bibfield  {author} {\bibinfo {author} {\bibfnamefont {T.}~\bibnamefont
  {{Bondo}}}, \bibinfo {author} {\bibfnamefont {M.}~\bibnamefont {{Hennrich}}},
  \bibinfo {author} {\bibfnamefont {T.}~\bibnamefont {{Legero}}}, \bibinfo
  {author} {\bibfnamefont {G.}~\bibnamefont {{Rempe}}}, \ and\ \bibinfo
  {author} {\bibfnamefont {A.}~\bibnamefont {{Kuhn}}},\ }\href
  {http://dx.doi.org/10.1016/j.optcom.2006.02.057} {\bibfield  {journal}
  {\bibinfo  {journal} {Opt. Commun.}\ }\textbf {\bibinfo {volume} {264}},\
  \bibinfo {pages} {271} (\bibinfo {year} {2006})}\BibitemShut {NoStop}%
\bibitem [{\citenamefont {Lindlein}\ \emph {et~al.}(2007)\citenamefont
  {Lindlein}, \citenamefont {Maiwald}, \citenamefont {Konermann}, \citenamefont
  {Sondermann}, \citenamefont {Peschel},\ and\ \citenamefont
  {Leuchs}}]{Lindl07}%
  \BibitemOpen
  \bibfield  {author} {\bibinfo {author} {\bibfnamefont {N.}~\bibnamefont
  {Lindlein}}, \bibinfo {author} {\bibfnamefont {R.}~\bibnamefont {Maiwald}},
  \bibinfo {author} {\bibfnamefont {H.}~\bibnamefont {Konermann}}, \bibinfo
  {author} {\bibfnamefont {M.}~\bibnamefont {Sondermann}}, \bibinfo {author}
  {\bibfnamefont {U.}~\bibnamefont {Peschel}}, \ and\ \bibinfo {author}
  {\bibfnamefont {G.}~\bibnamefont {Leuchs}},\ }\href
  {http://dx.doi.org/doi:10.1134/S1054660X07070055} {\bibfield  {journal}
  {\bibinfo  {journal} {Laser Phys.}\ }\textbf {\bibinfo {volume} {17}},\
  \bibinfo {pages} {927} (\bibinfo {year} {2007})}\BibitemShut {NoStop}%
\bibitem [{\citenamefont {Shu}\ \emph {et~al.}(2010)\citenamefont {Shu},
  \citenamefont {Kurz}, \citenamefont {Dietrich},\ and\ \citenamefont
  {Blinov}}]{Shu10}%
  \BibitemOpen
  \bibfield  {author} {\bibinfo {author} {\bibfnamefont {G.}~\bibnamefont
  {Shu}}, \bibinfo {author} {\bibfnamefont {N.}~\bibnamefont {Kurz}}, \bibinfo
  {author} {\bibfnamefont {M.~R.}\ \bibnamefont {Dietrich}}, \ and\ \bibinfo
  {author} {\bibfnamefont {B.~B.}\ \bibnamefont {Blinov}},\ }\href
  {http://dx.doi.org/10.1103/PhysRevA.81.042321} {\bibfield  {journal}
  {\bibinfo  {journal} {Phys. Rev. A}\ }\textbf {\bibinfo {volume} {81}},\
  \bibinfo {pages} {042321} (\bibinfo {year} {2010})}\BibitemShut {NoStop}%
\bibitem [{\citenamefont {Tey}\ \emph {et~al.}(2009)\citenamefont {Tey},
  \citenamefont {Maslennikov}, \citenamefont {Liew}, \citenamefont {Aljunid},
  \citenamefont {Huber}, \citenamefont {Chng}, \citenamefont {Chen},
  \citenamefont {Scarani},\ and\ \citenamefont {Kurtsiefer}}]{Tey09}%
  \BibitemOpen
  \bibfield  {author} {\bibinfo {author} {\bibfnamefont {M.~K.}\ \bibnamefont
  {Tey}}, \bibinfo {author} {\bibfnamefont {G.}~\bibnamefont {Maslennikov}},
  \bibinfo {author} {\bibfnamefont {T.~C.~H.}\ \bibnamefont {Liew}}, \bibinfo
  {author} {\bibfnamefont {S.~A.}\ \bibnamefont {Aljunid}}, \bibinfo {author}
  {\bibfnamefont {F.}~\bibnamefont {Huber}}, \bibinfo {author} {\bibfnamefont
  {B.}~\bibnamefont {Chng}}, \bibinfo {author} {\bibfnamefont {Z.}~\bibnamefont
  {Chen}}, \bibinfo {author} {\bibfnamefont {V.}~\bibnamefont {Scarani}}, \
  and\ \bibinfo {author} {\bibfnamefont {C.}~\bibnamefont {Kurtsiefer}},\
  }\href {http://dx.doi.org/10.1088/1367-2630/11/4/043011} {\bibfield
  {journal} {\bibinfo  {journal} {New J. Phys.}\ }\textbf {\bibinfo {volume}
  {11}},\ \bibinfo {pages} {043011} (\bibinfo {year} {2009})}\BibitemShut
  {NoStop}%
\bibitem [{\citenamefont {{Quinto-Su}}\ \emph {et~al.}(2004)\citenamefont
  {{Quinto-Su}}, \citenamefont {{Tscherneck}}, \citenamefont {{Holmes}},\ and\
  \citenamefont {{Bigelow}}}]{Quint04}%
  \BibitemOpen
  \bibfield  {author} {\bibinfo {author} {\bibfnamefont {P.~A.}\ \bibnamefont
  {{Quinto-Su}}}, \bibinfo {author} {\bibfnamefont {M.}~\bibnamefont
  {{Tscherneck}}}, \bibinfo {author} {\bibfnamefont {M.}~\bibnamefont
  {{Holmes}}}, \ and\ \bibinfo {author} {\bibfnamefont {N.~P.}\ \bibnamefont
  {{Bigelow}}},\ }\href {http://dx.doi.org/10.1364/OPEX.12.005098} {\bibfield
  {journal} {\bibinfo  {journal} {Opt. Express}\ }\textbf {\bibinfo {volume}
  {12}},\ \bibinfo {pages} {5098} (\bibinfo {year} {2004})}\BibitemShut
  {NoStop}%
\bibitem [{\citenamefont {{Wilzbach}}\ \emph {et~al.}(2009)\citenamefont
  {{Wilzbach}}, \citenamefont {{Heine}}, \citenamefont {{Groth}}, \citenamefont
  {{Liu}}, \citenamefont {{Raub}}, \citenamefont {{Hessmo}},\ and\
  \citenamefont {{Schmiedmayer}}}]{Wilzb09}%
  \BibitemOpen
  \bibfield  {author} {\bibinfo {author} {\bibfnamefont {M.}~\bibnamefont
  {{Wilzbach}}}, \bibinfo {author} {\bibfnamefont {D.}~\bibnamefont {{Heine}}},
  \bibinfo {author} {\bibfnamefont {S.}~\bibnamefont {{Groth}}}, \bibinfo
  {author} {\bibfnamefont {X.}~\bibnamefont {{Liu}}}, \bibinfo {author}
  {\bibfnamefont {T.}~\bibnamefont {{Raub}}}, \bibinfo {author} {\bibfnamefont
  {B.}~\bibnamefont {{Hessmo}}}, \ and\ \bibinfo {author} {\bibfnamefont
  {J.}~\bibnamefont {{Schmiedmayer}}},\ }\href
  {http://dx.doi.org/10.1364/OL.34.000259} {\bibfield  {journal} {\bibinfo
  {journal} {Opt. Lett.}\ }\textbf {\bibinfo {volume} {34}},\ \bibinfo {pages}
  {259} (\bibinfo {year} {2009})}\BibitemShut {NoStop}%
\bibitem [{\citenamefont {{Kohnen}}\ \emph {et~al.}(2011)\citenamefont
  {{Kohnen}}, \citenamefont {{Succo}}, \citenamefont {{Petrov}}, \citenamefont
  {{Nyman}}, \citenamefont {{Trupke}},\ and\ \citenamefont
  {{Hinds}}}]{Kohne09}%
  \BibitemOpen
  \bibfield  {author} {\bibinfo {author} {\bibfnamefont {M.}~\bibnamefont
  {{Kohnen}}}, \bibinfo {author} {\bibfnamefont {M.}~\bibnamefont {{Succo}}},
  \bibinfo {author} {\bibfnamefont {P.~G.}\ \bibnamefont {{Petrov}}}, \bibinfo
  {author} {\bibfnamefont {R.~A.}\ \bibnamefont {{Nyman}}}, \bibinfo {author}
  {\bibfnamefont {M.}~\bibnamefont {{Trupke}}}, \ and\ \bibinfo {author}
  {\bibfnamefont {E.~A.}\ \bibnamefont {{Hinds}}},\ }\href
  {http://dx.doi.org/10.1038/nphoton.2010.255} {\bibfield  {journal} {\bibinfo
  {journal} {Nat. Photonics}\ }\textbf {\bibinfo {volume} {5}},\ \bibinfo
  {pages} {35} (\bibinfo {year} {2011})}\BibitemShut {NoStop}%
\bibitem [{\citenamefont {{Vandevender}}\ \emph {et~al.}(2010)\citenamefont
  {{Vandevender}}, \citenamefont {{Colombe}}, \citenamefont {{Amini}},
  \citenamefont {{Leibfried}},\ and\ \citenamefont {{Wineland}}}]{VanDe10}%
  \BibitemOpen
  \bibfield  {author} {\bibinfo {author} {\bibfnamefont {A.~P.}\ \bibnamefont
  {{Vandevender}}}, \bibinfo {author} {\bibfnamefont {Y.}~\bibnamefont
  {{Colombe}}}, \bibinfo {author} {\bibfnamefont {J.}~\bibnamefont {{Amini}}},
  \bibinfo {author} {\bibfnamefont {D.}~\bibnamefont {{Leibfried}}}, \ and\
  \bibinfo {author} {\bibfnamefont {D.~J.}\ \bibnamefont {{Wineland}}},\ }\href
  {http://dx.doi.org/10.1103/PhysRevLett.105.023001} {\bibfield  {journal}
  {\bibinfo  {journal} {Phys. Rev. Lett.}\ }\textbf {\bibinfo {volume} {105}},\
  \bibinfo {pages} {023001} (\bibinfo {year} {2010})}\BibitemShut {NoStop}%
\bibitem [{\citenamefont {{Brady}}\ \emph {et~al.}()\citenamefont {{Brady}},
  \citenamefont {{Ellis}}, \citenamefont {{Moehring}}, \citenamefont {{Stick}},
  \citenamefont {{Highstrete}} \emph {et~al.}}]{Brady10}%
  \BibitemOpen
  \bibfield  {author} {\bibinfo {author} {\bibfnamefont {G.~R.}\ \bibnamefont
  {{Brady}}}, \bibinfo {author} {\bibfnamefont {A.~R.}\ \bibnamefont
  {{Ellis}}}, \bibinfo {author} {\bibfnamefont {D.~L.}\ \bibnamefont
  {{Moehring}}}, \bibinfo {author} {\bibfnamefont {D.}~\bibnamefont {{Stick}}},
  \bibinfo {author} {\bibfnamefont {C.}~\bibnamefont {{Highstrete}}},  \emph
  {et~al.},\ }\href {http://arxiv.org/abs/1008.2977} {\bibinfo  {journal}
  {arXiv:1008.2977}\ }\BibitemShut {NoStop}%
\bibitem [{\citenamefont {Schrama}\ \emph {et~al.}(1993)\citenamefont
  {Schrama}, \citenamefont {Peik}, \citenamefont {Smith},\ and\ \citenamefont
  {Walther}}]{Schra93}%
  \BibitemOpen
\bibfield  {journal} {  }\bibfield  {author} {\bibinfo {author} {\bibfnamefont
  {C.}~\bibnamefont {Schrama}}, \bibinfo {author} {\bibfnamefont
  {E.}~\bibnamefont {Peik}}, \bibinfo {author} {\bibfnamefont {W.}~\bibnamefont
  {Smith}}, \ and\ \bibinfo {author} {\bibfnamefont {H.}~\bibnamefont
  {Walther}},\ }\href {http://dx.doi.org/DOI: 10.1016/0030-4018(93)90318-Y}
  {\bibfield  {journal} {\bibinfo  {journal} {Opt. Commun.}\ }\textbf {\bibinfo
  {volume} {101}},\ \bibinfo {pages} {32 } (\bibinfo {year}
  {1993})}\BibitemShut {NoStop}%
\bibitem [{\citenamefont {Gulde}\ \emph {et~al.}(2001)\citenamefont {Gulde},
  \citenamefont {Rotter}, \citenamefont {Barton}, \citenamefont
  {Schmidt-Kaler}, \citenamefont {Blatt},\ and\ \citenamefont
  {Hogervorst}}]{Gulde01}%
  \BibitemOpen
  \bibfield  {author} {\bibinfo {author} {\bibfnamefont {S.}~\bibnamefont
  {Gulde}}, \bibinfo {author} {\bibfnamefont {D.}~\bibnamefont {Rotter}},
  \bibinfo {author} {\bibfnamefont {P.}~\bibnamefont {Barton}}, \bibinfo
  {author} {\bibfnamefont {F.}~\bibnamefont {Schmidt-Kaler}}, \bibinfo {author}
  {\bibfnamefont {R.}~\bibnamefont {Blatt}}, \ and\ \bibinfo {author}
  {\bibfnamefont {W.}~\bibnamefont {Hogervorst}},\ }\href
  {http://dx.doi.org/10.1007/s003400100749} {\bibfield  {journal} {\bibinfo
  {journal} {Appl. Phys. B}\ }\textbf {\bibinfo {volume} {73}},\ \bibinfo
  {pages} {861} (\bibinfo {year} {2001})}\BibitemShut {NoStop}%
\bibitem [{\citenamefont {Harlander}\ \emph {et~al.}(2010)\citenamefont
  {Harlander}, \citenamefont {Brownnutt}, \citenamefont {H\"ansel},\ and\
  \citenamefont {Blatt}}]{Harla10}%
  \BibitemOpen
  \bibfield  {author} {\bibinfo {author} {\bibfnamefont {M.}~\bibnamefont
  {Harlander}}, \bibinfo {author} {\bibfnamefont {M.}~\bibnamefont
  {Brownnutt}}, \bibinfo {author} {\bibfnamefont {W.}~\bibnamefont {H\"ansel}},
  \ and\ \bibinfo {author} {\bibfnamefont {R.}~\bibnamefont {Blatt}},\ }\href
  {http://dx.doi.org/10.1088/1367-2630/12/9/093035} {\bibfield  {journal}
  {\bibinfo  {journal} {New J. Phys.}\ }\textbf {\bibinfo {volume} {12}},\
  \bibinfo {pages} {093035} (\bibinfo {year} {2010})}\BibitemShut {NoStop}%
\bibitem [{\citenamefont {Kim}\ \emph {et~al.}(2010)\citenamefont {Kim},
  \citenamefont {Sushkov}, \citenamefont {Dalvit},\ and\ \citenamefont
  {Lamoreaux}}]{Kim10}%
  \BibitemOpen
  \bibfield  {author} {\bibinfo {author} {\bibfnamefont {W.~J.}\ \bibnamefont
  {Kim}}, \bibinfo {author} {\bibfnamefont {A.~O.}\ \bibnamefont {Sushkov}},
  \bibinfo {author} {\bibfnamefont {D.~A.~R.}\ \bibnamefont {Dalvit}}, \ and\
  \bibinfo {author} {\bibfnamefont {S.~K.}\ \bibnamefont {Lamoreaux}},\ }\href
  {http://dx.doi.org/10.1103/PhysRevA.81.022505} {\bibfield  {journal}
  {\bibinfo  {journal} {Phys. Rev. A}\ }\textbf {\bibinfo {volume} {81}},\
  \bibinfo {pages} {022505} (\bibinfo {year} {2010})}\BibitemShut {NoStop}%
\bibitem [{\citenamefont {Carmichael}\ and\ \citenamefont
  {Walls}(1976)}]{Carmi76}%
  \BibitemOpen
  \bibfield  {author} {\bibinfo {author} {\bibfnamefont {H.~J.}\ \bibnamefont
  {Carmichael}}\ and\ \bibinfo {author} {\bibfnamefont {D.~F.}\ \bibnamefont
  {Walls}},\ }\href {http://dx.doi.org/10.1088/0022-3700/9/8/007} {\bibfield
  {journal} {\bibinfo  {journal} {J. Phys. B}\ }\textbf {\bibinfo {volume}
  {9}},\ \bibinfo {pages} {1199} (\bibinfo {year} {1976})}\BibitemShut
  {NoStop}%
\bibitem [{\citenamefont {{Diedrich}}\ and\ \citenamefont
  {{Walther}}(1987)}]{Diedr87}%
  \BibitemOpen
  \bibfield  {author} {\bibinfo {author} {\bibfnamefont {F.}~\bibnamefont
  {{Diedrich}}}\ and\ \bibinfo {author} {\bibfnamefont {H.}~\bibnamefont
  {{Walther}}},\ }\href {http://dx.doi.org/10.1103/PhysRevLett.58.203}
  {\bibfield  {journal} {\bibinfo  {journal} {Phys. Rev. Lett.}\ }\textbf
  {\bibinfo {volume} {58}},\ \bibinfo {pages} {203} (\bibinfo {year}
  {1987})}\BibitemShut {NoStop}%
\bibitem [{\citenamefont {{Colombe}}\ \emph {et~al.}(2007)\citenamefont
  {{Colombe}}, \citenamefont {{Steinmetz}}, \citenamefont {{Dubois}},
  \citenamefont {{Linke}}, \citenamefont {{Hunger}},\ and\ \citenamefont
  {{Reichel}}}]{Colom07}%
  \BibitemOpen
  \bibfield  {author} {\bibinfo {author} {\bibfnamefont {Y.}~\bibnamefont
  {{Colombe}}}, \bibinfo {author} {\bibfnamefont {T.}~\bibnamefont
  {{Steinmetz}}}, \bibinfo {author} {\bibfnamefont {G.}~\bibnamefont
  {{Dubois}}}, \bibinfo {author} {\bibfnamefont {F.}~\bibnamefont {{Linke}}},
  \bibinfo {author} {\bibfnamefont {D.}~\bibnamefont {{Hunger}}}, \ and\
  \bibinfo {author} {\bibfnamefont {J.}~\bibnamefont {{Reichel}}},\ }\href
  {http://dx.doi.org/10.1038/nature06331} {\bibfield  {journal} {\bibinfo
  {journal} {\nat}\ }\textbf {\bibinfo {volume} {450}},\ \bibinfo {pages} {272}
  (\bibinfo {year} {2007})}\BibitemShut {NoStop}%
\bibitem [{\citenamefont {Sangouard}\ \emph {et~al.}(2009)\citenamefont
  {Sangouard}, \citenamefont {Dubessy},\ and\ \citenamefont {Simon}}]{Sango09}%
  \BibitemOpen
  \bibfield  {author} {\bibinfo {author} {\bibfnamefont {N.}~\bibnamefont
  {Sangouard}}, \bibinfo {author} {\bibfnamefont {R.}~\bibnamefont {Dubessy}},
  \ and\ \bibinfo {author} {\bibfnamefont {C.}~\bibnamefont {Simon}},\ }\href
  {http://dx.doi.org/10.1103/PhysRevA.79.042340} {\bibfield  {journal}
  {\bibinfo  {journal} {Phys. Rev. A}\ }\textbf {\bibinfo {volume} {79}},\
  \bibinfo {pages} {042340} (\bibinfo {year} {2009})}\BibitemShut {NoStop}%
\end{thebibliography}%
\end{document}